# Conductance spectra of (Nb, Pb, In)/NbP - superconductor/Weyl semimetal junctions

Revised….

G. Grabecki,[1] A. Dąbrowski,[1] P. Iwanowski,[1,2] A. Hruban,[1] B.J. Kowalski,[1] N. Olszowska,[3] J. Kołodziej,[3] M. Chojnacki,[1] K. Dybko,[1,2] A. Łusakowski,[1] T. Wojtowicz,[2] T. Wojciechowski,[1,2] R. Jakieła,[1] and A. Wiśniewski[1,2]

[1]*Institute of Physics, Polish Academy of Sciences, Aleja Lotnikow 32/46, PL-02668 Warsaw, Poland*
[2]*International Research Centere MagTop, Institute of Physics, Polish Academy of Sciences, Aleja Lotnikow 32/46, PL-02668 Warsaw, Poland*
[3]*National Synchrotron Radiation Centre SOLARIS, Jagiellonian University, Czerwone Maki 98, PL-30392 Kraków, Poland*

The possibility of inducing superconductivity in type-I Weyl semimetal through coupling its surface to a superconductor was investigated. A single crystal of NbP, grown by chemical vapor transport method, was carefully characterized by XRD, EDX, SEM, ARPES techniques and by electron transport measurements. The mobility spectrum of the carriers was determined and it was found that there are four separate sharp peaks visible, which indicates that the carriers participating in the conductance have four different (almost discrete) mobilities. For the studies of interface transmission, the (001) surface of the crystal was covered by several hundred nm thick metallic layers of either Pb, or Nb, or In. DC current-voltage characteristics and AC differential conductance through the interfaces as a function of the DC bias were investigated. Upon cooling of the devices during which the metals become superconducting, all three types of junctions show conductance increase, pointing out the Andreev reflection as a prevalent contribution to the subgap conductance. In the case of Pb-NbP and Nb-NbP junctions, the effect is satisfactorily described by modified Blonder-Tinkham-Klapwijk model. Unfortunately, the absolute value of the conductance is much smaller than that for the bulk crystal, indicating that the transmission occurs through only a small part of the contact area. An opposite situation occurs in In-NbP junction, where we observe very high and narrow peak at zero bias. The conductance at the peak reaches the bulk value indicating that almost whole contact area is transmitting and, additionally, a superconducting proximity phase is formed in the material. We interpret this as a result of indium diffusion into NbP, where the metal atoms penetrate the surface barrier and form very transparent superconductor-Weyl semimetal contact inside. However, further diffusion occurring already at room temperature leads to degradation of the effect, so it is observed only in the pristine structures. Despite of this, our observation directly demonstrates possibility of inducing superconductivity in a type-I Weyl semimetal.

# I. INTRODUCTION

There is currently a broad interest in materials of topologically nontrivial band structure and their studies have become the most active field of the condensed matter research [1-3]. The essence of one class of these materials, called topological insulators, is an existence of topologically protected metallic surface states having Dirac-like dispersion, while their bulk is gapped and thus not conducting. Another class are Weyl semimetals (WSM) whose bulk conduction and valence bands show conical dispersions forming (one or multiple) *pairs of Weyl nodes* in the momentum space [4,5]. Each of the nodes of the pair has opposite chirality. This differentiates WSMs from the usual Dirac semimetals [6,7] where the analogical nodes do not show chirality and can be treated as two degenerate Weyl nodes, shifted to the same point in the **k**-space. The most important characteristic of WSMs is that the nodes of opposite chirality give rise to formation of discontinuous Fermi surface manifested as Fermi arcs beginning and ending at the projection of the bulk Fermi surface in reciprocal space [8]. It has to be also stressed that WSMs are possible only in systems without center of inversion or when the time reversal symmetry is broken. It was shown that depending on whether the system conserves or violates Lorentz symmetry, WSMs can be divided into two classes: type-I and type-II. Photoemission spectroscopy measurements have shown that NbP and TaAs are type-I WSMs and that WTe$_2$ belongs to the type-IIclass. The type-I WSMs can be considered as a direct negative band gap semiconductors while the type-II ones are characterized by an indirect negative gap. [9].

Recently, an idea of introducing superconductivity in topological materials has been developed owing to the possibility of non-zero momentum Cooper pairing [10-14]. One of the recent spectacular demonstration in this area was experimental observation of the Klein tunneling through PtIr/SmB$_6$/YB$_6$ structure, when superconductivity was induced in the topological insulator SmB$_6$ due to proximity with YB$_6$[15]. The tunneling produced 100% transmission despite of interface barriers whose existence was proved by the Authors. Another consequence of the non-zero momentum pairing is the possibility of formation of zero-energy modes that are equivalent to Majorana fermions [16, 17]. These fermions, governed by non-Abelian statistics, show potential for practical realization of fault-tolerant topological quantum computation [18-20]. Although there is still a very long way to build such quantum computer, one needs to find an optimal material platform for its construction. In particular, introducing nonzero superconducting order parameter into topological materials by inducing superconductivity through the proximity effect enables to employ presently achievable WSMs and conventional superconductors. This is believed to be much easier than synthetizing new intrinsic materials, with both WSM and superconducting characteristics.

Interestingly, most of the experimental research on superconductor-WSM (S-WSM) structures performed so far, have been done for type-II Weyl semimetals. Kononov et al. [13] studied charge transport through the interface between a niobium superconductor and WTe$_2$ which is a typical type-II WSM. They observed signature of Fermi arc surface states in Andreev reflection. In further studies of electron transport between two superconducting indium leads coupled to a single crystal of WTe$_2$, the same research group [21] demonstrated Josephson current in In-WTe$_2$-In junctions, confirmed by the observation of integer and fractional Shapiro steps under microwave irradiation. In recent work [22], another group have shown that surface steps on WTe$_2$ layers led to an occurrence of one–dimensional superconductivity.

The experimental papers devoted to type-I WSMs are much more scarce. It is well known that proximity effect at the superconductor – normal metal (S-N) junction strongly depends on the interface quality [23]. In particular, it is significantly limited by interface barriers arising due to surface contamination during the structure fabrication. An interesting approach to overcome this problem was presented in the paper of Maja Bachmann *et al.* [24], wherein the authors studied a related WSM, niobium arsenide (NbAs). They irradiated the samples by Ga$^+$ ion beam which depleted 20 nm-deep surface layer of As atoms, forming Nb-NbAs junction buried inside the material. Performed electron transport measurements showed superconductivity in the surface layer and clear indications of its penetration into the nearby bulk NbAs. However, a drawback of this method was significantly reduced $T_c$ in the Nb layer, which was explained in terms of contamination by residual As atoms. Additionally, Maja Bachmann *et al.* did not study conductance perpendicular to the S-WSM junction and thus were not able to evaluate the possibly existing interface barrier.

In view of the present interest in WSMs, it is pertinent to perform experimental studies of conductance across S-WSM junctions prepared in the conventional way, i.e. by deposition of the superconductor on the crystal. Although this method carries the risk of interface contamination, the possible reduction of the transmissivity may be balanced by better technological flexibility and well controlled properties of the superconducting layer. Additionally, for deposited S-WSM junctions produced by S deposition direct measurements of the interface conductance can be easily performed. A goal of the present research is to study conductance across S-WSM interface for another compound of the transition metal monopnictides, namely niobium phosphide (NbP) type-I WSM. Similarly to NbAs, NbP is also noncentrosymmetric WSM containing 12 pairs of Weyl nodes in the Brillouin zone [25]. The NbP single crystals were obtained by chemical vapor transport (CVT) method. Crystals used for superconductors deposition were first examined by X-ray diffraction (XRD), energy-dispersive X–ray spectroscopy (EDX), scanning electron microscopy (SEM), angle-resolved photoelectron spectroscopy (ARPES) techniques and by electron transport measurements (resistivity and Hall). As superconductors, we used lead, niobium and indium whose thin layers were deposited on (001) oriented surfaces. We verified that all three metallic layers show good superconducting properties when cooled down to helium temperatures. In order to evaluate the interface transmission, we measured DC current-voltage characteristics and AC differential conductance through the interfaces. The measurements were carried out as a function of temperature and magnetic field and the results were interpreted in the frame of the modified Blonder-Tinkham Klapwijk (BTK) model [23,26]. The model describes the interface barrier in terms of a single parameter $Z_{eff}$ and takes into account finite lifetime of the transmitted quasiparticles $\tau = \hbar/\Gamma$.

When each of the three junctions is cooled so the metal becomes superconducting, pronounced conductance rises in the subgap regions are observed. In the case of Pb-NbP junction, the perpendicular differential conductance as a function of the bias voltage increases up to 40 %, indicating considerable contribution of the Andreev reflection. It is well described by modified BTK theory with $Z_{eff}$ = 0.47 and $\Gamma$ = 0.10 meV, for all studied temperatures. On the other hand, the absolute value of the interface conductance indicates that the effective area of the junction is only a minute part of the area of the deposited Pb layer. For Nb-NbP junction, the effective area is also reduced but is about 10 times larger than for Pb-NbP case. Here, the subgap conductance increase is even higher, reaching 82% at zero voltage bias. The fit to modified BTK theory gives the barrier parameter, $Z_{eff}$ = 0.4 and $\Gamma$ = 0.03 meV. However, for Nb-NbP junction the description is less accurate than for the Pb-NbP, because of current-induced superconductivity suppression in Nb layer. Much interesting behavior is observed in In-NbP structures, where the subgap conductance increases by much more than 2 times in comparison to the normal state. Additionally, the differential conductance shows very large peak at zero bias whose maximum value is almost equal to the conductance of the bulk NbP and thus indicates negligible contact resistance. The peak disappears at voltages and magnetic fields much smaller than those for the In layer which points to occurrence of some superconducting phase in the material due to the proximity effect. Unfortunately, the effect is not permanent because of significant In diffusion into NbP occurring during keeping the sample at room temperature.

Our results show that deposition of all three metals have potential for exploring proximity induced effects in NbP. Further improvement of the interface quality with respect of its effective area and durability would enable fabrication of more sophisticated S-WSM low-dimensional structures aimed at searching for Majorana fermions.

## II. EXPERIMENTAL DETAILS

### A. Synthesis of polycrystalline NbP

In synthesis of Weyl semimetal NbP we took into consideration physical and chemical properties of its constituting elements (Nb, P). Niobium has a high melting point (2470 $^{o}$C) and chemical inertness while phosphorus is chemically active and easily sublimates at low temperature. NbP compound, similarly to TaAs, decomposes before reaching its melting point [27].

In our synthesis processes, we applied fully controlled and safe procedure to avoid quartz ampule explosion and to produce stoichiometric polycrystalline NbP to be used as a source charge for successive single crystal growths. The conditions in experimental set up for synthesis process are presented in Fig. 1. It was achieved in the two zone horizontal furnace with temperature in synthesis area ~800 $^{o}$C and in phosphorus source area ~460 $^{o}$C. In a typical run, Nb 10 μm powder (Alfa Aesar – 99.99%) compressed into ø10 mm tablets and red phosphorus (Heraens – 99.999%) were placed in ø20 mm, l = 220 mm quartz ampule sealed under vacuum $10^{-6}$ hPa. Synthesis took place at temperatures 800 $^{o}$C – 850 $^{o}$C. The temperature of cold end of the quartz ampule set to 460 $^{o}$C guaranteed safe and stable phosphorus vapour pressure ~1000 hPa. The process lasting ~80 h resulted in producing of black polycrystalline NbP.

### B. NbP monocrystals growth

For the growth of NbP crystal we applied Chemical Vapor Transport (CVT) method. The experimental heating system, the same as used for synthesis of polycrystalline NbP consisted of two-zone furnace (shown in Fig. 1.).

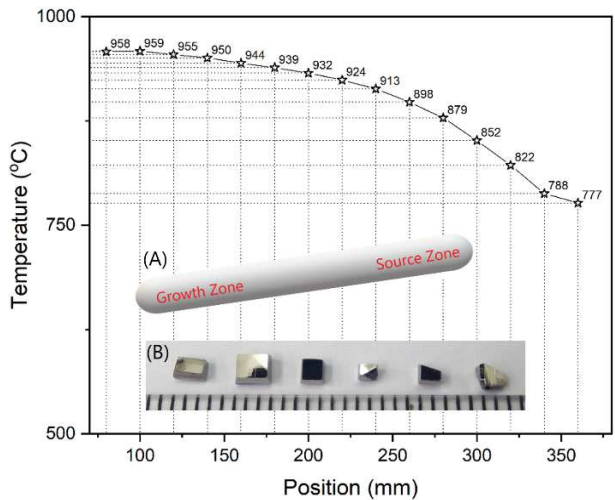

Fig. 1. Temperature gradients in heating system, insets: (A) schematic representaion of the ampule; (B) NbP single crystals that were grown (lower tics distance is 1 mm)

Starting micro polycrystalline NbP material compressed into the form of tablets was loaded into the quartz ampule (inner ø16 mm, l = 210 mm) together with $J_2$ transporter (6-10 mg/cm$^3$) and some amount of red phosphorus. After charging the evacuated and sealed ampule was placed into the furnace and gradually heated to the conditions, with the growth temperature gradient from 950 $^o$C (growth zone) to 850 $^o$C (source zone). The CVT process was run for two weeks and then the furnace was cooled down to room temperature with 100 $^o$C/h cooling rate. Typical NbP crystals that were grown are presented in the inset to Fig. 1.

### C. Structural characterization by XRD and SEM/EDX

Crystal structure and crystallographic quality of grown NbP were verified by X-ray powder diffraction using Rigaku SmartLab 3kW diffractometer equipped with a tube having Cu anode, and operating with U = 40kV and I = 30mA. The characteristic peak positions were identified based on the data from ICDD PDF-4+ 2018 RDB database (DB card number: 04-003-0878) – see Fig. 2. All measured samples were oriented along the <001> direction. The single crystal orientation was performed on a KUMA-diffraction four-circle diffractometer with a goniometer in kappa geometry and the peak-hunting procedure was used. An elementary cell, cell constants, and orientations of the main crystallographic axes were determined. NbP crystallizes in a body-centered tetragonal unit cell and its space group is I41md (No. 109). The lattice constants determined for our crystal were: a = 3.33544(6) Å, c = 11.3782(3) Å and V = 126.584(4) Å$^3$.

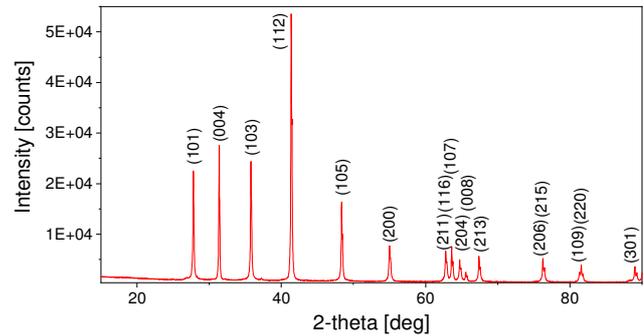

Fig. 2. Experimental XRD diffraction pattern of grown NbP crystal.

The quantitative chemical composition of NbP was studied using energy dispersive X ray spectroscopy (EDX) system QUANTAX 400 Bruker coupled with the Zeiss Auriga field emission (Schottky type) scanning electron microscope (FE-SEM), operating at 15 kV incident energy. The measurements were performed with the use of oriented samples and their results confirmed that NbP crystals are well stoichiometric (within experimental error the atomic ratio was 1:1).

### D. ARPES measurements

ARPES experiments were carried out with use of the UARPES beamline at the National Synchrotron Radiation Centre SOLARIS in Krakow (Poland). Elliptically polarizing quasiperiodic undulator of APPLE II type was used as the source of the radiation in the energy range of 8–100 eV. The end-station was equipped with the SCIENTA OMICRON DA30L photoelectron spectrometer. The high energy and angular resolution of the spectrometer (1.8 meV and 0.1$^o$, respectively) enabled the band mapping with a precision sufficient for observations of the features of the band structure characteristic of Weyl semimetals.

### E. Electron transport measurements

For electron transport measurements, we prepared a rectangular sample of dimensions 2 mm×1 mm×0.5 mm. Six gold wires, 50 μm thick, were attached to the sides of the sample in Hall bridge geometry using small drops of silver paste. The resistance and Hall effect measurements were performed in the temperature range from 1.5 K to 200 K and in the magnetic fields up to 6 T. We applied phase-sensitive AC technique (using Stanford Research SR830 lock-in) with AC electric current (f = 19 Hz) in the miliampere range.

### F. Deposition of superconducting layers on NbP surface

In order to perform studies of S-WSM interfaces, various thin metallic layers were evaporated onto the (001) NbP surface. The first one was a pure niobium layer deposited by a magnetron sputtering technique using a target with a purity of 99.995%. The NbP monocrystalline substrate sample was thoroughly cleaned by rinsing in acetone, trichloroethylene, ethanol and isopropyl alcohol. The deposition process was performed through a mask with 0.9 mm diameter circular opening, in a UHV chamber with the base pressure of $4.3 \cdot 10^{-9}$ hPa. The NbP crystal was preheated to the temperature of about 150°C in order to facilitate desorption of atmospheric gases from the surface. The deposition rate of niobium was 3.6 nm/min and the final thickness of Nb thin film was 149 nm.

In the case of Pb, a layer in the form of a circular spot with a diameter of 1 mm and thickness of 240 nm was deposited on the NbP crystal through the metal mask having the circular aperture. Prior to the loading of molybdenum block with crystal that was kept by the mask into the Molecular Beam Epitaxial (MBE) system, the crystal was thoroughly cleaned in hot Trichloroethylene and methanol. Pb flux was supplied from effusion cell equipped with PBN crucible filled with 6N pure Pb. During the deposition of Pb, that took place at the rate of 6 nm/min, the NbP was kept at 20 C and the pressure inside the chamber was around $1 \times 10^{-10}$ hPa.

Finally, indium layers were thermally evaporated in Balzers' evaporator with a glass chamber and visual control of the process. The substrate was cooled by liquid nitrogen. The pressure during the process was approx. $2 \times 10^{-7}$ hPa. The deposited layer thickness was estimated from the geometry of the system and from the mass of the evaporated metal. Its surface was metallically gleaming and the thickness was 300 ±100 nm.

### G. S-WSM interface conductance measurements

In order to measure the interfacial transmission, thin gold wires were attached to the metal layers and NbP substrate using small drops of silver paste. We verified that these contacts did not introduce any traces of superconductivity in the entire measured temperature range. The sample geometry is illustrated in the left inset to Fig. 8(A). Interface conductance was measured with the use of three-probe method which allowed to avoid contribution from Ag-NbP interface. Double Au wires have been attached to each contact. This enabled us to evaluate the Ag-NbP contact resistance without contribution of the resistance of the measurement lines. For DC measurements we used Keithley 6882 current source and Keithley 199 digital multimeter. AC measurements were always performed during the same cooling sessions as DC measurements. Because the interface resistances were smaller than resistance of the wires connecting the sample to the measurement system, we swept DC current $I$, instead of voltage $U$. The current was modulated by 100 μA AC component of frequency 19 Hz. The resulting AC voltage signal was measured by SR830 lock-in nanovoltmeter. All measurements were performed in a home-made cryostat equipped with 9 T superconducting coil. The sample was installed in variable-temperature insert, where the temperature was controlled with the accuracy of couple of mK.

## III. EXPERIMENTAL RESULTS AND DISCUSSION

### A. ARPES studies

ARPES studies confirmed that grown NbP crystals are high quality Weyl semimetals. Figure 3 shows the results for NbP single crystal cleaved along the (001) plane under UHV condition (the base pressure of $5 \times 10^{-11}$ hPa). The band structure scans of NbP were taken at the temperature of 78 K for the photon energies of 93.4 eV (Fig. 3(A)-(C)) and 44.4 eV (Fig. 3(D)). The radiation linear polarization was kept horizontal. The total energy resolution during the data acquisition (limited both by the energy resolution of the electron energy analyzer and the spectral width of the synchrotron radiation beam) was 10 meV and the angular resolution was $0.1^0$.

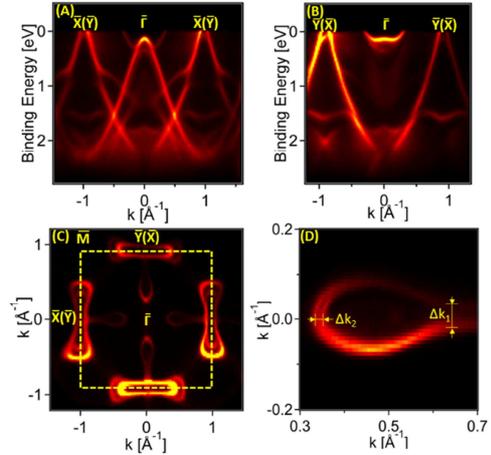

Fig. 3. The experimental band structure cuts taken by ARPES for the NbP(001) surface at the temperature of 78 K for the photon energy of 93.4 eV: (A) along the $\bar{X}$-$\bar{\Gamma}$-$\bar{X}$ direction of the surface Brillouin zone (for the dominating surface configuration – see the text), (B) along the $\bar{Y}$-$\bar{\Gamma}$-$\bar{Y}$ direction, (C) constant energy cut at the Fermi energy (the surface Brillouin zone boundaries marked with the broken line). (D) (data taken for $hv$ = 44.4 eV) shows one of the "spoon-like" Fermi surface pockets close to the Γ point.

Figures 3(A)-(C) show three perpendicular cross sections of the 2D band structure of NbP plotted parallel to the (001) plane (the sample surface): A - along the $\bar{X}$-$\bar{\Gamma}$-$\bar{X}$ direction, B - along $\bar{Y}$-$\bar{\Gamma}$-$\bar{Y}$, C – the constant energy plot at the Fermi energy. In the latter case, the Fermi surface pockets terminated by Fermi arcs, the fingerprints of Weyl semimetal character of the investigated system, manifest themselves clearly in the vicinity of the $\bar{\Gamma}$ point (the "spoon-like" ones) or close to the $\bar{X}$ point (the "bowtie-like" ones). This diagram is composed of two rotated by $90^0$ overlapping contributions from surface domains with Nb-P broken bonds aligned along the mutually perpendicular directions. As one of the contributions is markedly stronger, the description of the BZ points corresponding to the weaker one is given in parenthesis. The overall shape of the band structure derived from these results corresponds well to the ARPES data and *ab initio* calculation for NbP reported in literature [25]. Figure 3(D) shows details of the "spoon-like" Fermi surface pocket occurring close to the $\Gamma$ point. It is possible to resolve the splitting of this feature caused by the spin-orbit interaction. The values for this splitting in k-space assessed from our data ($\Delta k_1 = 0.051$ Å$^{-1}$, $\Delta k_2 = 0.016$ Å$^{-1}$) are somewhat higher than those estimated in [25], however they still follow the trend determined for NbP, TaP and TaAs as a function of increasing spin-orbit coupling.

### B. Electron transport measurements

Further characterization of NbP single crystals has been made by electron transport measurements using Hall bridge configuration. Figure 4(A) presents magnetoresistance measured at different temperatures, in the range from 1.8 K to 200 K. At the lowest temperature, magnetoresistance is approx. 900 %, and drop down by about 3 times at 200 K. The Hall resistivity vs. magnetic field *B* presented in Fig. 4(B), shows nonlinear behavior with the slope increasing with increasing *B*. Its maximal slope at *T* =1.8 K reaches 0.81 cm$^3$/C and it shows n-type conductivity. As the temperature is increased, the slope consistently decreases and finally the conductance changes to p-type.

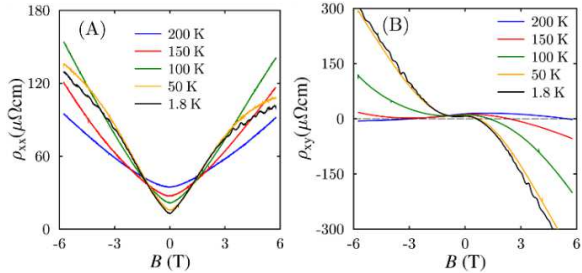

Fig. 4. (A) Longitudinal magnetoresistivity of NbP crystal measured for various temperatures. (B) The same for Hall resistivity.

The lowest-temperature magnetoresistance shows pronounced Shubnikov-de Haas (SdH) oscillations which are plotted in Fig. 5(A) after subtracting monotonic background. In the case of single periodicity, the minima should be periodic in $1/B$ scale, however we observe much more complex picture indicative of several contributing periodicities. In order to resolve them, we performed Fast Fourier Transform (FFT) analysis whose results are represented in Fig. 5(B). It revealed that the complex SdH oscillation picture is a result of superposition of four frequencies, resulting from contributions of 4 different Fermi surface cross sections perpendicular to the magnetic field direction.

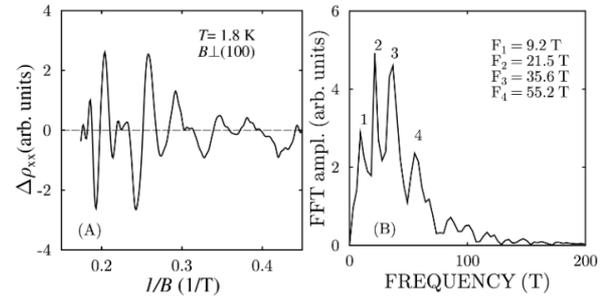

Fig. 5. (A) Shubnikov-de Haas oscillations in NbP single crystal measured for the magnetic field perpendicular to the (100) direction. (B) Fast Fourier spectrum of the oscillations shown in Fig. 5 A.

Due to a complicated multi-band character of conductivity in NbP, it is difficult to directly obtain concentrations and mobility of carriers in different bands. For this reason, we performed mobility spectrum analysis [28-31]. This method relies on numerical fitting of the mobility dependent conductivity density $S(\mu)$ to experimentally obtained conductivity tensor components $\sigma_{xx}(B)$ and $\sigma_{xy}(B)$. The mobility spectrum $S(\mu)$ usually consists of distinct peaks which correspond to different kinds of carriers with broadening of each peak indicating energy dependence of relaxation time. We applied such mobility spectrum analysis according to a numerical recipe described in Refs. [29, 30]. The results are shown in Fig. 6. The most important observation is the occurrence of four separate sharp peaks indicating that the carriers participating in the conductance have 4 different, and almost discrete mobilities. It has to be noted here that negative values are for holes and positive for electrons. At $T = 1.8$ K, the peak positions indicate that the mobility values are equal to: $\mu_{h1} = -3.7$, $\mu_{h2} = -1.3$, $\mu_{e1} = 0.94$ and $\mu_{e2} = 5.6$ in units of $10^4$ cm$^2$/Vs. The peaks from both the electrons and hole sides shift towards zero as the temperature is increased, which is a result of a mobility decrease stemming from phonon scattering. In particular, at $T = 200$ K, we obtain respective values: $\mu_{h1} = -0.82$, $\mu_{h2} = -0.063$, $\mu_{e1} = 0.44$ and $\mu_{e2} = 1.6$ (all in $10^4$ cm$^2$/Vs).

Whether these four mobility peaks correspond to the four Fermi surface cross sections revealed by the SdH FFT analysis is not clear at this stage of research.

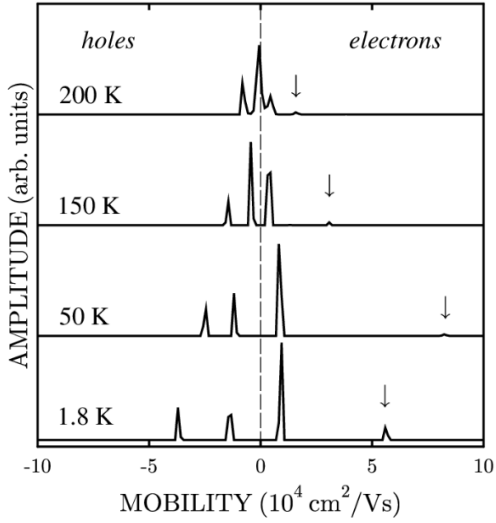

Fig. 6. Mobility spectra for NbP single crystal measured in magnetic field **B** ⊥ (100), at various temperatures.

Summarizing the electron transport characterization, we have to stress that our NbP crystals show *qualitatively* similar properties to those for which results already been published in literature [32-35]. In particular, the resistance tensors $\rho_{xx}(B)$ and $\rho_{xy}(B)$ (Fig. 4(A), (B)) show very similar characteristics to those presented in Refs. [32, 33]. Furthermore, the FFT analysis performed in Ref. [32] gives spectra corresponding well to these shown in Fig. 5(B). However, there are also important *quantitative* differences. First of all, the resistance ratio $\rho_{xx}(200\ K)/\rho_{xx}(1.8\ K) \approx 3$, is lower than the previously reported. Additionally, the resistance of our crystal at 1.8 K increases by about 9 times when B is raised to 6 T, while the reported values of positive magnetoresistance are of the order of 1000. Such a large unsaturated magnetoresistance is a result of mixed electron and hole conductivity of the material [34]. Also, strongly nonlinear dependence of Hall resistance on the magnetic field indicates mixed n-type/p-type character of the conductivity. The electron mobility at 1.8 K for the most mobile carriers reaches only the value of around 6 m$^2$/Vs, which is also much smaller than previously reported values exceeding 100 m$^2$/Vs [32,33]. One possible reason for these differences might be a much higher contribution of electrons to electrical transport in our crystals. In particular, $\rho_{xy}(B)$ measured at 1.8 K shows the high field slope of about 0.82 cm$^3$/C. This is 5 times smaller than analogical result given in Ref. [32] and indeed points to much higher value of Fermi energy in our crystals. This may indicate a presence of excess donor with high concentration.

### C. Conductance through metal-NbP interface

Our principal results are conductance measurements across S-WSM interfaces performed as a function of the bias voltage. They correspond to sweeping the Fermi energy across the energy gap of the superconductor, $E_g(T)=2\Delta(T)$, and provide information on the subgap transport. In particular, for transparent interfaces in the voltage range $U(T) \leq \pm \Delta(T)/e$, an increase by the factor of 2 is expected because the conduction occurs by two-particle Andreev reflection process. In the opposite case of opaque interface, the subgap conductance drops down to zero and the case of Giaver tunneling occurs [36]. The entire range of transparencies between these two limits was described by the theory formulated by Blonder, Tingham and Klapwijk [23]. This theory assumed δ-Dirac interface barrier described by a single parameter $Z_{eff}$. A current flowing through the interface may be described as:

$$I = C \int_{-\infty}^{\infty}[f(E - eU) - f(E)][1 + A(E) - B(E)]dE. \quad (1)$$

Here, $f(E)$ is the Fermi-Dirac distribution, $A(E)$ – probability of Andreev reflection and $B(E)$ – a coefficient of usual reflection, $C$ is a constant dependent on the interface area and density of states. BTK model was later improved by introducing finite lifetime of the quasi particles penetrating the barrier, $\tau = \hbar/\Gamma$ [26], and we used this version to interpret our results. It has to be noted that even in the case of a lack of any interface barrier ($Z=0$), the transmission is usually lower than 100%, because of Fermi velocity mismatch between the superconducting layer and the normal metal (in our case being WSM). This can be described in terms of an effective barrier $Z_{eff}$ [37]:

$$Z_{eff}^2 = Z^2 + \frac{(1-r)^2}{4r}, \quad (2)$$

where $r \equiv \frac{v_{FN}}{v_{FS}}$, is the ratio of Fermi velocities in the normal metal (N) and the superconductor (S). For this reason we use $Z_{eff}$ instead of $Z$ in our calculations. Additionally, Eq. (2) allows us also to estimate the Fermi velocity in the NbP crystal.

To apply BTK theory, we need to know parameters of the superconducting layers deposited on the NbP crystals, which may be different from the tabularized values for pure bulk superconductors. Therefore, we measured the critical temperatures and calculated values of $\Delta(0)$ using standard formula:

$$2\Delta(0) = 3.52 k_B T_c. \quad (3)$$

For non-zero temperatures, the superconducting energy gap can be approximated by the empirical formula [38]:

$$\Delta(T) = \Delta(0) \tanh\left(1.74 \times \sqrt{\frac{T_c}{T} - 1}\right), \quad (4)$$

which gives precision better than 5 %. Finally, measurements in the magnetic field were performed to determine the dependence of the critical field on temperature:

$$B_c(T) = B_c(0)\left(1 - \frac{T^2}{T_c^2}\right), \quad (5)$$

which would be another proof of the quality of the superconducting layers.

1. **Pb-NbP interface**

Superconducting properties of the deposited layers were studied with the use of the same samples that were used for studies of the interface conductance. As it is shown in Fig. 7(A) and (B), the critical temperature for our Pb layers is almost the same as for bulk Pb, $T_c$ = 7.2 K. Also the critical magnetic field is increased only by 30 % with respect to the bulk value. DC current-voltage characteristic of Pb-NbP junction is presented in Fig. 8(A). For $T < T_c$, we observe pronounced slope change in the middle part which corresponds to the conductance increase in the subgap range. The details are better visible in AC differential conductance $G = dI/dU$ shown in Fig. 8(B). The data are normalized to the value at high biases, $G_\infty$ = 0.218 $\Omega^{-1}$. At $T$ = 1.8 K, differential conductance increases by almost 40%, which indicates significant contribution of the Andreev reflection mechanism, but there is also some interface scattering present. To quantify it, we employ the modified BTK model with the energy gap parameter $\Delta(T)$ obtained from Eqs. (3) and (4). The fitted curves (solid) for various temperatures are anchored at the experimental points for $U$ = 0. They all correspond to the same barrier $Z_{eff}$ = 0.47 and lifetime $\Gamma$ = 0.10 parameters. It has to be noted that $G_\infty$ is much smaller than the NbP bulk conductance (see Fig. 4(A)). For the present NbP structure we have $R < 1$ m$\Omega$, corresponding to $G = 1/R > 10^3$ $\Omega^{-1}$. Most probably, this indicates that the effective contact area of the Pb-NbP junction is only $10^{-3}$-$10^{-4}$ part of its geometrical area. This is presumably a result of surface contamination (oxidation) during the fabrication process. Nevertheless, our observation of good transmission properties even for such minute surface indicates that further technological improvement is feasible.

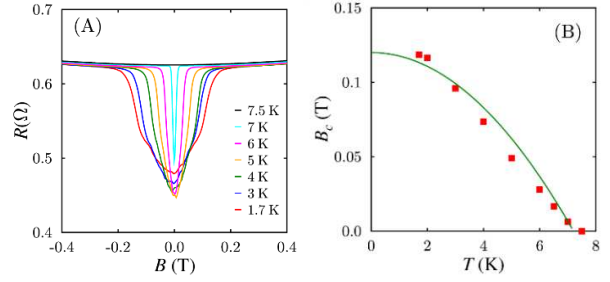

Fig. 7. (A) Interface resistance of Pb-NbP for $I$ = 100 µA measured as a function of the magnetic field, at various temperatures. (B) Measured critical field of the Pb layer as a function of temperature. Green line represents fit of Eq. (5) to the experimental data.

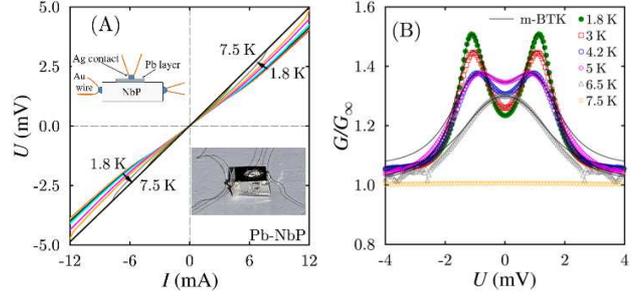

Fig. 8. (A) Voltage-current characteristics of Pb-NbP junction measured at different temperatures between 1.8 K and 7.5 K (with step of approximately 1 K), at zero magnetic field. The left inset represents sample scheme with three contacts and double Au wires attached. The right inset represents a photo of the sample with superconducting island on top and attached wires. (B) Differential conductance as a function of voltage across Pb-NbP junction measured at various temperatures (symbols), $G_\infty$ = 0.218 $\Omega^{-1}$. Solid lines are fits of modified BTK theory with $Z_{eff}$ = 0.45 and $\Gamma$ = 0.13 meV for T=1.8 K, and $Z_{eff}$ = 0.47 and $\Gamma$ = 0.10 meV for other temperatures.

Using the value of $Z_{eff}$, obtained from the BTK theory fitting, one can estimate the range of possible Fermi velocities in NbP crystal. We consider two extreme cases for two terms in Eq. (2): i) $Z = 0$ and $Z_{eff}^2 = \frac{(1-r)^2}{4r}$ (there is no interface barrier and the scattering occurs only due to Fermi velocity mismatch), ii) $Z_{eff} = Z = 0.47$ and $r = 1$ (Fermi velocities are equal and the scattering is only due to the barrier). Taking known value of the Fermi velocity in bulk Pb, $v_{F-Pb}$ = 1.83×10$^6$ m/s [39], we obtained the range of possible values of $v_{F-NbP}$ between 0.74×10$^6$ m/s and 4.54×10$^6$ m/s.

2. **Nb-NbP interface**

For the Nb layer (Fig. 9(A), (B)), we found $T_c$ = 7.5 K, which is distinctly smaller than standard bulk value $T_c$ = 9.3 K. Additionally, the critical magnetic field $B_c$ reaches values as high as 3 T, which is probably a result of II-type character of the layer superconductivity [36].

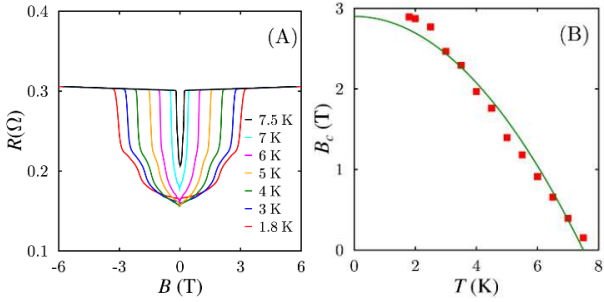

Fig. 9. (A) Interface resistance of Nb-NbP junction measured with current of 100 μA, as a function of the magnetic field at various temperatures. (B) Critical field $B_c$ of the Nb layer as a function of temperature (points). Green solid curve represents a fit of formula (5).

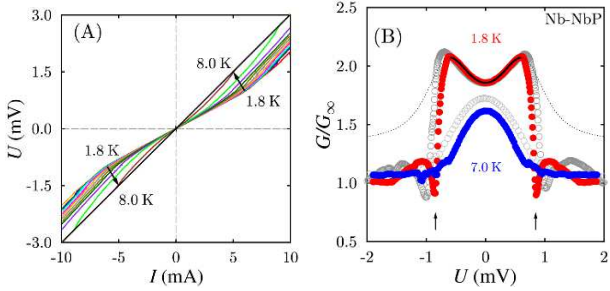

Fig. 10. (A) U-I characteristics of Nb-NbP junction measured at various temperatures in the range from 1.8 K to 8 K (with a step of about 0.5 K), and at zero magnetic field. (B) Differential conductance of Nb-NbP junction measured at 1.8 K and 7 K ($G_\infty$ = 2.8 Ω$^{-1}$). Filled symbols represent data obtained just after fabrication of the sample, empty symbols denote the data obtained on the same junction after 20 months of storing it at room temperature. Solid/dotted curve is calculated with modified BTK model for T = 1.8 K, assuming $\Delta(0)$ = 0.7 meV. The fit is valid only for the central part (solid line) and not valid outside (dotted line) because of the current-induced S-N transition (indicated by arrows). Obtained barrier parameter is $Z_{eff}$ = 0.4 and energy broadening is $\Gamma$ = 0.03 meV

DC voltage-current characteristics measured at different temperatures are shown in Fig. 10(A). Above the superconducting transition, at T=8 K, we observe perfect ohmic behavior. For T<$T_c$, pronounced slope decrease occurs in the middle voltage range, U ≤ ±0.8 mV. At 1.8 K, the slope reduction is close to 2 times when compared to that at 8 K, indicating considerable subgap transmission.

More details are derived from differential conductance, G measured as a function of voltage and shown in Fig. 10(B). For clarity, we present only the data for two temperatures 1.8 K and 7 K. Similarly to the Pb-NbP case, the absolute conductance value, $G_\infty$ = 2.8 Ω$^{-1}$, is about 10$^3$ times smaller than bulk conductance of NbP crystal of comparable dimensions (see Fig 4(A)). Therefore, the active part of the junction area is a minute part of its geometrical value equal to 0.6 mm$^2$. At 1.8 K, the subgap conductance increase is very abrupt and, additionally, two sharp dips occur at its edges, at $U_d \approx \pm 0.84$ mV, where the conductance is smaller as compared to the normal state. Such dips have been observed in other S-N junctions [40], and they were explained in terms of the current-induced transition to the normal state in the superconductor layer. Assuming that the active contact area contains only 10$^{-3}$ of the geometrical one, we expect current densities as high as 10$^6$ A/cm$^2$ at the voltage corresponding to the dip positions. This is just the order of magnitude characteristic for critical currents in thin layers of Nb [41]. Although the BTK model does not account for critical current-induced transition to the normal state, it well describes the 1.8 K data for |U| < $U_d$. For this range, the fit to the modified BTK theory (thick solid line), gives $Z_{eff}$ = 0.40, $\Gamma$=0.03 meV, and $\Delta(0)$ = 0.7 meV. The theory is unable to describe the curve for T = 7 K, because of the increasing thermal voltage broadening ~kT/e, which contributes to the BTK model (see Eq. (1)). As it is seen from Fig. 10(B), $G/G_\infty$ ratio at 1.8 K reaches values even larger than 2, the maximal increase possible in BTK model. This may be related to the over-critical current flowing in Nb layer at |U| > $U_d$ which produces Joule heat locally warming the junction and lowering the conductance [40, 42]. Finally, we have to mention that the studied Nb-NbP junction is very stable. The conductance spectrum was almost the same when the measurements were repeated after 20 months from junction fabrication (see empty symbols in Fig. 10(B)).

In analogy to the previously discussed Pb-NbP junction, we substituted $Z_{eff}$ = 0.40 into Eq. (2) to estimate the mean Fermi velocity in NbP, $v_{F\text{-}NbP}$. Again, we considered two extreme cases: i) Z = 0 and $Z_{eff}^2 = \frac{(1-r)^2}{4r}$; ii) $Z_{eff} = Z$ = 0.40 and r = 1. Inserting known value of Fermi velocity in bulk Nb, $v_{F\text{-}Nb}$ = 1.37×10$^6$ m/s [39], we obtained the possible interval for $v_{F\text{-}NbP}$ values between 0.55×10$^6$ m/s and 3.4×10$^6$ m/s . Comparing this estimation with that obtained for Pb-NbP junction, we obtain possible range of Fermi velocity in NbP: 0.74×10$^6$ m/s < $v_{F\text{-}NbP}$ < 3.4×10$^6$ m/s. This range is quite wide, however, it encompasses Fermi velocities known for other materials with Dirac dispersions, such as graphene [43]. This fully confirms WSM nature of NbP which contains multiple bands with linear dispersion.

### 3. In-NbP interface

The superconducting transition in the In layer deposited on NbP takes place at $T_c \approx 3.8$ K, which is somewhat higher than that for bulk In, which equals 3.4 K. Additionally, the superconductivity persists in magnetic fields up to 500 mT, much stronger than bulk critical values (Fig. 11).

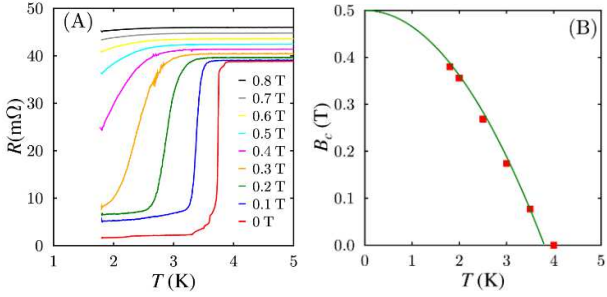

Fig. 11. (A) Resistance of In-NbP junction measured as a function of temperature at different magnetic fields. (B) Critical magnetic fields of the In-NbP structure as a function of temperature. Green solid curve represents a fit of formula (5) to the experimental data.

For this junction, we observed conductance spectrum qualitatively different from those observed for the previous two. The data obtained just after the fabrication are shown in Fig 12 (A), (B). At high $U$ $G_\infty \approx 26$ $\Omega^{-1}$, which is already much larger than the corresponding $G$ values for Pb-NbP and Nb-NbP contacts. For lower $|U|$, $G$ exhibits unexpected asymmetry with respect to $U = 0$. In both DC and AC measurements, we observe abrupt conductance increase occurring at about -0.3 mV and +0.5 mV. The asymmetric conductance plateau is formed around $U = 0$, where, at $T = 2$ K, $G \approx 170$ $\Omega^{-1}$ on negative, and $G \approx 130$ $\Omega^{-1}$ at positive side, respectively. When the temperature is increased, the plateau becomes narrower and the asymmetry tends to vanish. In the right inset to Fig.12 (B) we show magnified central region for $U \leq \pm 75$ μV. There is a conductance peak in the range between -30 μV and +50 μV. The peak itself is quite symmetric, but its sides evolve into the two asymmetric plateau at different values of $U$ and $G$. At the peak maximum, about $U \approx \pm 10$ μV, we see a pair of symmetric dips located on both sides of the sharp central maximum, where $G$ peaks up to almost 800 $\Omega^{-1}$. This part is magnified in the left inset to Fig 12 (B). The central maximum shows some asymmetry, but this asymmetry is uncertain because of finite accuracy of the $U$ scale in the sub-microvolt range marked by the horizontal bar.

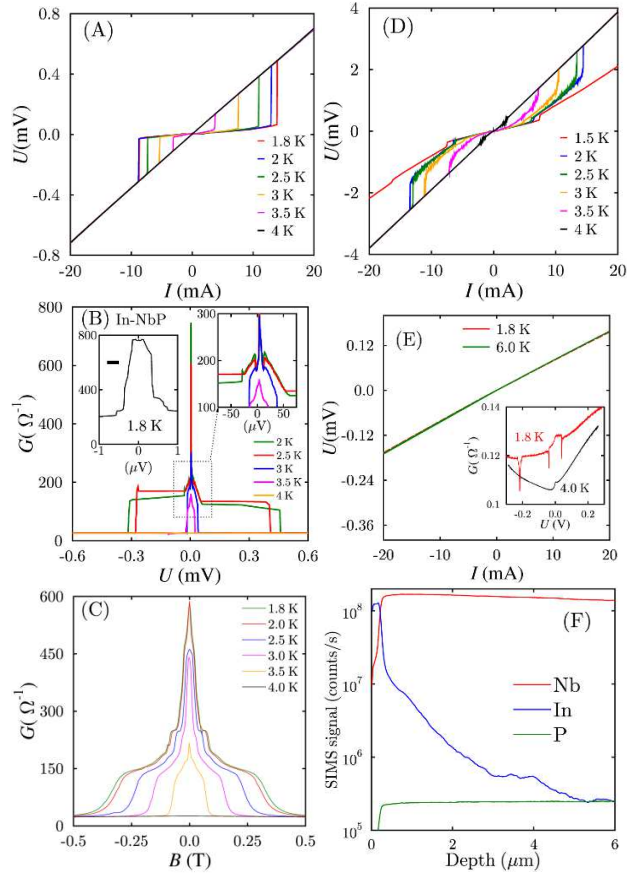

Fig. 12. (A) $U$-$I$ characteristics of the In-NbP junction measured at different temperatures, just after its fabrication. (B) Differential conductance $G = dI/dU$ for different temperatures measured just after device fabrication. (Left inset): contains expanded region around the central peak measured at $T = 1.8$ K. Horizontal bar represents error of the horizontal axis. (Right inset): Magnified central region, indicated by the dotted rectangle. (C) Conductance at $U = 0$, measured as a function of the magnetic field, at different temperatures. (D) $U$-$I$ characteristics of the same junction measured after its 6 months long storage at room temperature. (E) $U$-$I$ characteristics measured after 16 months long room temperature storage. Inset represents differential conductance $G$ collected for the same structure, at the same time. (F) SIMS depth profile of the junction taken after 16 months long room temperature storage.

Application of magnetic field leads to consecutive suppression of all the conductance features discussed above. This is illustrated by the conductance data for $U = 0$, presented in Fig. 12(C). Here, all the curves are symmetric with respect to $B = 0$. At the lowest temperature, there is an abrupt decrease in the range $\pm 30$ mT, where $G$ decreases from 600 $\Omega^{-1}$ to about 250 $\Omega^{-1}$. This corresponds to a suppression of the conductance at the central narrow peak.

The next decrease, from 250 Ω⁻¹ to about 150 Ω⁻¹, occurring at ±80 mT, well corresponds to suppression of the intermediate peak in the range between -30 μV and +50 μV. Finally, the two slopes around ±340 mT correspond to vanishing of the wide flat plateau.

Unfortunately, when the sample is stored at room temperature, the transport properties gradually change with time. This is illustrated in Fig. 12 (D) and (E). These figures show DC *I-V* curves, measured after 6 and 16 months from the fabrication time, respectively. Already in the former case the changes are substantial and all asymmetry virtually disappeared. In the latter case the curves become almost linear. Then AC differential conductance (Fig. 12 (E) inset) shows only residual variation, in the range 0.1-0.14 Ω⁻¹, which suggest that complete deterioration of the In-NbP junction took place. In order to determine whether the origin of this process is In diffusion, we have studied the chemical profile by means of secondary ion mass spectroscopy (SIMS). The measurements were carried out 16 months after the In deposition and the result is shown in Fig. 12(F). One can clearly see constant signals from Nb and P atoms of the host crystal and the variable signal from In that extends up to 6 μm into the crystal. Given the fact that we have observed progressive changes in the *I-V* curves with time, from the moment of junction formation trough 6$^{th}$ month until 16$^{th}$ month, we believe that the diffusion was taking place during entire period of the sample storage at room temperature. We note that similar indium diffusion occurring at room temperature was also observed in other crystals, like lead telluride PbTe [44, 45].

Very high conductance values observed for the pristine junction indicate that the In-NbP contact's active area is comparable to the geometrical one, in contrast to Pb-NbP and Nb-NbP junctions. Such a radical reduction of possible surface barriers could be achieved due to indium diffusion into NbP. Shortly after the fabrication, In diffusing atoms may surmount surface barriers and produce highly transparent S-WSM interface within the crystal [46]. In analogy to the results presented in Ref. [24], such interface would be free of contamination. For this reason, the observations done on the pristine In-NbP junctions may give us an access to physical effects occurring at the very S-WSM interface, independently of the later fate of the samples stored at room temperature. This is impossible in the case of the Pb-NbP and Nb-NbP junctions, because their minute effective contact surface area precludes such access. Possibly, this is also a reason that their conductance spectra can be described by the BTK model that does not accounts for any effects resulting from unusual structure of the Weyl semimetal surface [8]. In contrast, pristine In-NbP junction shows conductance spectrum which is very complicated and cannot be described by the BTK model. In particular, the most central peak in *G(U)* dependence, which occurs within $U < \pm 1 \mu V$ and is suppressed at the magnetic field that are much smaller than the In layer's critical field, points to formation of some new superconducting proximity phase in the NbP [47]. The intermediate peak occurring in the range (-30 μV, +50 μV) rises above the background from 130 Ω⁻¹ to 220 Ω⁻¹, i.e. conductance increases about 1.7 times. This is consistent with the Andreev reflection across S-WSM interface, however the energy gap value is 10 times smaller than that in the bulk In [48]. Finally, the two conductance jumps occurring at -0.3 mV and +0.5 mV (both ends of the asymmetric conductance plateau in Fig. 12(B)) are possibly related to the critical current in the superconducting layer. However, the asymmetry suggests that the effect cannot be related to superconductivity of In layer, but rather to some interaction with the underlying WSM, for which various nonreciprocal responses are expected [49]. Explanation of all these unusual effects awaits a proper theoretical model.

## IV. CONCLUSIONS

We studied conductance through the superconductor–type-I Weyl semimetal interface for three different metals deposited on the (100) surface of NbP monocrystals. The most important observation was done on In-NbP junction, where near-zero bias conductance peak indicated highly transparent contact area. It results from In diffusion into the substrate which gives rise to formation of highly transparent S-WSM interface, free of surface contamination. This high interface transmission occurs only in the pristine junctions, right after their fabrication. In this situation, occurrence of superconducting proximity phase in the material is possible, whose clear indications have been observed. Although further diffusion leads to disappearance of the proximity phase, our result indicates that its formation in type-I Weyl semimetals is possible. In the case of the two other studied junctions, Pb-NbP and Nb-NbP the interface conductance values are much smaller indicating strongly reduced active junction areas. However, their differential conductance spectra as a function of voltage show pronounced increases in the subgap regions. This indicates good transmissions in these small active regions. The results for Pb-NbP and Nb-NbP junctions are well described by BTK theory with moderate barrier parameters $Z_{eff} = 0.47$ and $Z_{eff} = 0.40$, respectively. In the case of Nb-NbP junction, the model description is valid only in the limited range of bias voltages because of an abrupt transition of the Nb metal layer into normal state caused by the current flowing through the small active contact area. Long-term stability of the Nb-NbP junction indicates lack of the metal diffusion into NbP crystal, which is in contrast to the In-NbP case. Therefore, the metal diffusion is an important factor in achieving highly transparent S-WSM interfaces.

## ACKNOWLEDGEMENTS

The research was partially supported by the Foundation for Polish Science through the IRA Program co-financed by the European Union within SGOP.

1. C. L. Kane and E. J. Mele, Phys. Rev. Lett. **95**, 146802 (2005).
2. M. Z. Hasan and C. L. Kane, Rev. Mod. Phys. **82**, 3045 (2010).
3. D. A. Pesin and L. Balents, Nature Phys. **6**, 376 (2010).
4. H. Weyl, Phys. **56**, 330 (1929).
5. N. P. Armitage, E.J. Mele, and A. Vishwanath, Rev. Mod. Phys. **90**, 015001 (2018).
6. S. M. Young, S. Zaheer, J. C. Y. Teo, C. L. Kane, E. J. Mele, and A. M. Rappe, Phys. Rev. Lett. **108**, 140405 (2012).
7. S. Borisenko, Q. Gibson, D. Evtushinsky, V. Zabolotnyy, B. Buchner, R. J. Cava, Phys. Rev. Lett. **113**, 027603 (2014).
8. M. Z. Hasan, S. Y. Xu, I. Belopolski, and S. M. Huang, Ann. Rev. Condens. Matter Phys. **8**, 289 (2017).
9. Hao Zheng and M. Zahid Hasan, Adv. in Physics: X, 3:1, 1466661 (2018).
10. B. A. Bernervig and T. Hughes, *Topological Insulators and Topological Superconductors*, (Princeton University Press, Princeton, New Jersey, 2013).
11. Y.Qi, P. G. Naumov, M.N. Ali, C. R. Rajamathi, W. Schnelle, O. Barkalov, M. Hanfland, S.-C. Wu, C. Shekhar, Y. Sun, V. Süß, M. Schmidt, U. Schwarz, E. Pippel, P. Werner, R. Hillebrand, T. Förster, E. Kampert, S. Parkin, R. J. Cava, C. Felser, B. Yan and S. A. Medvedev, Nat. Commun. **7**:10038 doi: 10.1038/ncomms11038 (2016).
12. O. O. Shvetsov, V. A. Kostarev, A. Kononov, V. A. Golyashov, K. A. Kokh, O. E. Tereshchenko and E. V. Deviatov, EPL 119, 57009 (2017).
13. A. Kononov, O. O. Shvetsov, S. V. Egorov, A. V. Timonina, N.N. Kolesnikov and E. V. Deviatov, EPL, **122**, 27004 (2018).
14. P. Schüffelgen, D. Rosenbach, E. Neumann, M. P. Stehno, M. Lanius, J. Zhao, M. Wang, B. Sheehan, M. Schmidt, B. Gao, A. Brinkman, G. Mussler, T. Schäpers, D. Grützmacher, J. Cryst. Growth **477**, 183–187 (2017).
15. S. Lee, V. Stanev, X. Zhang, D. Stasak, J. Flowers, J.S. Higgins, S. Dai, T. Blum, X. Pan, V. M. Yakovenko, J. Paglione, R. L. Greene, V. Galitski, and Ichiro Takeuchi, Nature **570**, 344 (2019).
16. E. Majorana, Nuovo Cim. **14**, 171–184 (1937).
17. F. Wilczek, Nat. Phys. **5**, 614–618 (2009).
18. J. Alicea, Rep. Prog. Phys. **75**, 076501 (2012).
19. T. Hyart, B. van Heck, I. C. Fulga, M. Burrello, A. R. Akhmerov, and C. W. J. Beenakker, Phys. Rev. **B** 88, 035121 (2013).
20. B. Lian, X.-Q. Sun, A. Vaezi, X.-L. Qi, and S.-C. Zhang, P. Natl. Acad. Sci. (USA) **115**, 10938 (2018).
21. O. O. Shvetsov, A. Kononov, A. V. Timonina, N. N. Kolesnikov and E. V. Deviatov EPL, **124**, 47003 (2018).
22. A. Kononov, G. Abulizi, K. Qu, J. Yan, D. Mandrus, K. Watanabe, T. Taniguchi, and C. Schnenberger, arXiv: 1911.02414v1.
23. G. E. Blonder, M. Tinkham and T. M. Klapwjik, Phys. Rev. B **25**, 4515 (1982).
24. M. D. Bachmann, N. Nair, F. Flicker, R. Ilan, T. Meng, N. J. Ghimire, E. D. Bauer, F. Ronning, J. G. Analytis, and P. J. W. Moll, Sci. Adv. **3**: e1602983 (2017).
25. Z. K. Liu, L. X. Yang, Y. Sun, T. Zhang, H. Peng, H. F. Yang, C. Chen, Y. Zhang, Y. F. Guo, D. Prabhakaran, M. Schmidt, Z. Hussain, S.-K. Mo, C. Felser, B. Yan, Y. L. Chen, Nat. Mat. **15**, 27 (2015) and Suppl. Information.
26. A. Plecenik, M. Grajcar, S. Benacka, P. Seidel and A. Pfuch, Phys. Rev. B **49**, 10016 (1994).
27. Z. Li, H. Chen, S. Jin, D. Gan, W. Wang, L. Guo, and X. Chen, Cryst. Growth Des., **16**(3), 1172 (2016)
28. W. A. Beck and J. R. Anderson, J. Appl. Phys. **62**, 541 (1987).
29. I. Vurgaftman, J. R. Meyer, C. A. Hoffman, D. Redfern, J. Antoszewski, L. Faraone, and J. R. Lindemuth, J. Appl. Phys. **84**, 4966 (1998).
30. D. Chrastina, J.P. Hague, D.R. Leadley, J. Appl. Phys. **94**, 6583 (2003).
31. S. Kiatgamolchai, M. Myronov, O. A. Mironov, V. G. Kantser, E. H. C. Parker, and T. E. Whall, Phys. Rev. E **66**, 036705 (2002).
32. C. Shekhar, A. K. Nayak, Y. Sun, M. Schmidt, M. Nicklas, I. Leermakers, U. Zeitler, Y. Skourski, J. Wosnitza, Z. K. Liu, Y. L. Chen, W. Schnelle, H. Borrmann, Y. Grin, C. Felser, and B. H. Yan, Nat. Phys. **11**, 645 (2015).
33. P. Kumar, P. Neha, T. Das, and S. Patnaik, Sci. Rep. 7, 46062 (2017).

34. A. Leahy, Y.-P. Lin, P. E. Siegfried, A. C. Treglia, J. C. W. Song, R. M. Nandkishore, and M. Lee, Proc. Natl. Acad. Sci. (USA) **115**, 10570 (2018).
35. P. Sergelius, J. Gooth, S. Bäßler, R. Zierold, Ch. Wiegand, A. Niemann, H. Reith, C. Shekhar, C. Felser, B. Yan and K. Nielsch, *Sci. Rep.* **6**, 33859; doi: 10.1038/srep33859 (2016).
36. See e.g. M. Tinkham, *Introduction to Superconductivity*, (Courier Corporation, North Chelmsford MA, 2004).
37. G. E. Blonder and M. Tinkham, Phys. Rev. B **27**, 112 (1983).
38. N. Yabuki, *et al.* Nat. Commun. **7**:10616 doi: 10.1038/ncomms10616 (2016).
39. N.W. Ascroft, N. D. Mermin, *Solid State Physics*, (Holt, Rinehart and Winston, New York 1976).
40. G. Sheet, S. Mukhopadhyay, and P. Raychaudhuri, Phys. Rev. **B** 69, 134507 (2004).
41. R. P. Huebener, R. T. Kampwirth, R. L. Martini, T. W. Barbee, Jr. and R. B. Zubeckt, IEEE Transactions on Magnetics, MAG-11, no. 2, 344 (1975).
42. R. Haussler, G. Goll, Yu. G. Naidyuk, and H. v. Lohneysen, Physica B **218**, 197 (1996).
43. Ch. Hwang, D. A. Siegel, S-K. Mo, W. Regan, A. Ismach, Y. Zhang, A. Zettl and Al. Lanzara, Sci. Rep. 2:590.
44. B. Chang, K. E. Singer, and D. C. Northrop, J. Phys. D: Appl. Phys. **13**, 715 (1980).
45. G. Grabecki, K. A. Kolwas, J. Wróbel, K. Kapcia, R. Puźniak, R. Jakieła, M. Aleszkiewicz, T. Dietl, G. Springholz, and G. Bauer, J. Appl. Phys. **108**, 053714 (2010).
46. I. E. Batov, Th. Schäpers, A. A. Golubov, and A. V. Ustinov, J. Appl. Phys. **96**, 3366 (2004).
47. P. G. De Gennes, *Supreconductivity of Metals and Alloys*, (CRC Press, Taylor & Francis Group, Boca Raton 1999).
48. W.D. Gregory, L.S. Straus, R.F. Averill, J.C. Keister, C. Chapman, *Bulk Tunneling Measurements of the Superconducting Energy Gaps of Gallium, Indium, and Aluminum*, in: K.D. Timmerhaus, W.J. O'Sullivan, E.F. Hammel (eds) Low Temperature Physics-LT 13. (Springer, Boston, MA 1974)
49. Y. Tokura and N. Nagaosa, Nat. Commun. 9, 3740 (2018); A. Johansson, J. Henk, and I. Mertig, Phys. Rev. B **97**, 085417 (2018).